\documentstyle[twoside,fleqn,espcrc2,epsf]{article}

\newcommand{\lsim}{\rlap{\raise 2pt \hbox{$<$}}{\lower 2pt \hbox{$\sim$}}}
\newcommand{\gsim}{\rlap{\raise 2pt \hbox{$>$}}{\lower 2pt \hbox{$\sim$}}}
\newcommand{\etal}{{\it et al.}}

\title{Atmospheric muons and neutrinos from charm\thanks{Presented by
Paolo Gondolo.}}

\author{Paolo Gondolo\address{Universit\'e de Paris VII, Physique
Th\'eorique et Hautes Energies, 2
place Jussieu, 75005 Paris, France}\address{University of Oxford,
Department of Physics, 1 Keble Road, Oxford, OX1
3NP, United Kingdom},
Gunnar Ingelman\address{Department of Radiation Sciences,  Uppsala University,
 Box 535, S-751 21 Uppsala, Sweden }\address{Deutsches 
Elektronen-Synchrotron DESY, 
Notkestrasse 85, D-22603 Hamburg, Germany}
and Mats Thunman$^{\rm c}$}
       
\begin{document}

\begin{abstract}
We have updated our previous investigation of the production of muons
and neutrinos in cosmic ray interactions with the atmosphere, taking
account of recent results from the $ep$ collider HERA in our QCD-based
model for hadronic interactions. Qualitatively, our previous results
remain unmodified: our predictions for the conventional muon and
neutrino fluxes agree with earlier calculations, whereas the charm
particle treatment we use gives significantly lower prompt fluxes
compared to earlier estimates. This implies better prospects for
detecting very high energy neutrinos from cosmic sources.
\end{abstract}

\maketitle

The flux of muons and neutrinos at the Earth has an important
contribution from decays of particles produced through the interaction
of cosmic rays in the atmosphere.  This has an interest in its own
right, since it reflects primary interactions at energies that can by
far exceed the highest available accelerator energies. It is also a
background in studies of neutrinos from cosmic sources as attempted in
large neutrino telescopes.  

We have updated our detailed study of muon and neutrino production in
cosmic ray interactions with nuclei in the atmosphere
\cite{thunman,stockholm} by taking into account the information on
nucleon structure functions from recent experiments at the $ep$
collider HERA \cite{hera1,hera2} and by refining our simulation of
particle cascades in the atmosphere with a better treatment of
secondary particles.  A complete description is presented in
ref.~\cite{GIT}.

Our investigations agree with earlier studies on the fluxes of
atmospheric muons and neutrinos coming from decays of $\pi$ and $K$
mesons. On the other hand, we find quite low fluxes for prompt muons
and neutrinos, which arise through semi-leptonic decays of hadrons
containing heavy quarks (most notably charm). Other estimates of these
prompt fluxes
\cite{Volkova87,Castagnoli84,Gonzalez,Zas93,Bugaev,other-refs} are
higher than ours, and vary by few orders of magnitude. These
differences come from extrapolating charm production data from
accelerator energies to the orders-of-magnitude higher energies of the
relevant cosmic ray collisions. Current data from surface and
underground detectors attempting to measure the flux of prompt muons
and neutrinos (see {\it e.g.\/}~\cite{Bugaev-at-nestor}) are still too
discrepant to discriminate between the different models for charm
production.

The main contribution of our study is in using proper charm production
data and a sound physical model based on QCD.  First, we use
recent charm cross section measurements that form a consistent set of
data, but disagree with some of the early measurements that were
substantially higher. Secondly, we apply state-of-the-art models to
simulate charm particle production in high energy hadron-hadron
interactions.

We have obtained the atmospheric muon and neutrino fluxes with two
different methods: via a Monte Carlo simulation of the hadronic
cascade and via approximate analytical expressions with
energy-dependent $Z$-moments.  We were satisfied that the two methods,
which are conceptually rather different, gave quite similar
results. Differences were typically less than 20\%, below the
uncertainty in our charm calculation and quite acceptable in this
context.

\begin{figure*}[hbt]
\epsfxsize=160mm 
\centerline{\epsfbox{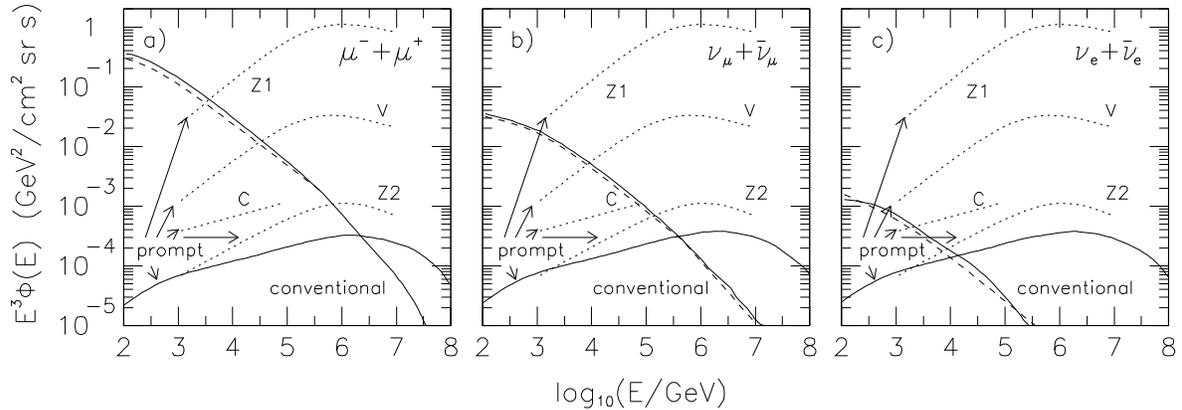}} 
\vspace{-2pc}
\caption{Our prompt and conventional
muon and neutrino fluxes (solid lines) compared with those in other
models for charm production.}
\label{fig:8}
\end{figure*}

Very different prompt fluxes are predicted by other models for charm
production, which differ both in the magnitude and energy dependence
of the total charm production cross section and in the distribution in
longitudinal momentum fraction of the charmed particles.

\begin{figure}[htb]
\epsfxsize=80mm 
\centerline{\epsfbox{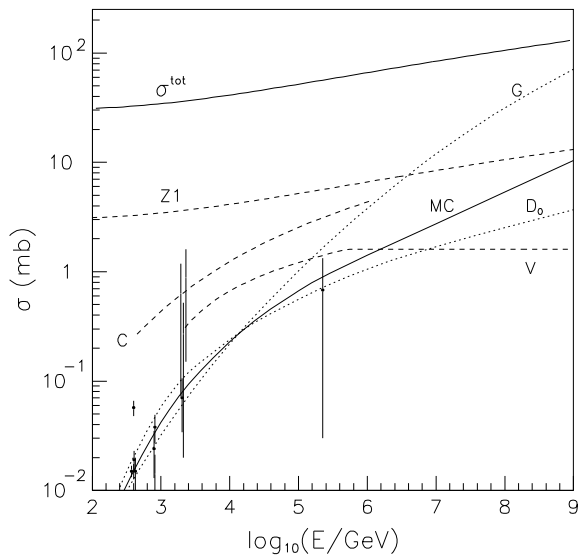}}
\vspace{-2pc}
\caption{Energy dependence of charm production cross section in pp
(p$\bar{\rm p}$) collisions. References to experimental points are
given in ref.~\protect\cite{GIT}. Curves are described in the text.}
\label{fig:cross}
\vspace{8pt}
\vspace{-\baselineskip}
\end{figure}

Volkova \etal~\cite{Volkova87}, applying the so-called `quark-gluon
string model' \cite{qgsm}, obtained the curves labeled V in
fig.~\ref{fig:8}. Castagnoli \etal~\cite{Castagnoli84} obtained the
result marked C. Both models use a parametrized energy dependent charm
cross section (curves C and V respectively in fig.~\ref{fig:cross})
normalized to early experimental data which are substantially above
later measurements. A more recent calculation based on the QGSM by
Gonzalez-Garcia \etal~\cite{Gonzalez} gives fluxes that are comparable
to curve C in fig.~\ref{fig:8}.

Curves marked Z1 and Z2 in figs.~\ref{fig:8} and~\ref{fig:cross} are
from Zas \etal~\cite{Zas93}. Curves Z1 illustrate an extreme model
where the charm cross section is simply taken as 10\% of the total
inelastic cross section (which is curve $\sigma^{\rm tot}$ in
fig.~\ref{fig:cross}). This is substantially higher than all charm
data. Curves Z2 correspond to charm {\em quark} production calculated
with leading order perturbative QCD matrix elements using relatively
hard parton distributions.

All the previous models except Z2 assume Feynman scaling for the charm
energy spectra. Bugaev \etal~\cite{Bugaev} considered Feynman scaling
violations through a phenomenological equation.  Their calculation
resulted in overall prompt fluxes slightly lower than Volkova's (curve
V). Their fluxes are higher in the non-scaling case than in the
scaling case (opposite to our results).

Our model (solid lines in fig.~\ref{fig:8} and curve MC in
fig.~\ref{fig:cross}) uses leading order perturbative QCD matrix
elements supplemented by a correction $K=2$ for next-to-leading order
processes. It incorporates a strong breaking of Feynman scaling,
arising from the dominance of perturbative charm quark production
close to the charm threshold.  For the parton densities in the nucleon
we have adopted the $MRS\,G$ parametrization \cite{mrsg}, which uses
essentially all relevant experimental data, from deep inelastic
scattering experiments to recent results from the $ep$ collider HERA
\cite{hera1,hera2}. Actually, since a naive extrapolation of the $G$
parametrization below the measured region in $x$ at rather small $Q^2
\sim m_c^2$ leads to a presumably unphysically large charm production
cross section, we have implemented a flatter dependence as $x\!\to\!0$
like $x^{-\epsilon}$ with $\epsilon \simeq 0.08$
(cfr.~\cite{SS}). Curves labeled G and D$_0$ in fig.~\ref{fig:cross}
show the charm production cross section one would obtain from the
(unflattened) $G$ parametrization and from the now-unacceptable
$MRS\,D_0$ parametrization \cite{mrsd0}.

Some of the previous models give charm production cross sections
higher than recent data (see fig.~\ref{fig:cross}) and some assume Feynman
scaling for the longitudinal momentum distributions. In some cases one
may have been mislead in the construction of the models by the early
charm measurements that turned out to be substantially higher than the
measurements done later.  Our model instead gives a fair description
of measured charm production cross sections and applies well-motivated
charm particle momentum distributions with significant Feynman scaling
violations.

With respect to other models for charm production, we predict
substantially lower muon and neutrino atmospheric fluxes from decays of
charm mesons. In particular, we predict an interestingly low prompt
neutrino background to searching high energy neutrino cosmic sources
in large scale neutrino telescopes.

\vfill\eject

\end{document}